\documentclass[12pt]{article}
\usepackage{epsf}
\begin{document}

\title{Spectral Density of Complex Networks \\ with a Finite Mean Degree}

\author{Taro Nagao$^1$ and G.J. Rodgers$^2$}

\date{}
\maketitle

\begin{center}
\it
$^1$ Graduate School of Mathematics,
Nagoya University, Chikusa-ku, \\ Nagoya 464-8602, Japan \\
\it
$^2$ Department of Mathematical Sciences, Brunel University, 
Uxbridge, Middlesex, UB8 3PH, United Kingdom

\end{center}

\begin{abstract}

In order to clarify the statistical features of complex networks, 
the spectral density of adjacency matrices has often been investigated. 
Adopting a static model introduced by Goh, Kahng and Kim, we analyse the 
spectral density of complex scale free networks. For that purpose, we 
utilize the replica method and effective medium approximation 
(EMA) in statistical mechanics. As a result, we identify a new integral 
equation which determines the asymptotic spectral density of scale 
free networks with a finite mean degree $p$. In the limit $p 
\rightarrow \infty$, known asymptotic formulae are rederived. 
Moreover, the $1/p$ corrections to known results are analytically 
calculated by a perturbative method. 
   
\end{abstract}


\medskip


\newpage

\section{Introduction}
\setcounter{equation}{0}
\renewcommand{\theequation}{1.\arabic{equation}}

The theory of complex networks, which originates in 
Erd\"os and R\'enyi's study on random graphs\cite{ER}, 
has dramatically developed in the last decade. Complex 
networks in biological and social systems were extensively 
investigated and universal features were elucidated. 
Moreover, as Erd\"os and R\'enyi's random graphs could 
not reproduce some of those features, new theoretical 
models of complex networks were introduced\cite{WS,BA}. 
\par        
One of the newly discovered features of complex 
networks is the scale free property, which means 
that the degree (the number of the nodes directly 
connected to each node) distribution obeys a power law. 
The scale free property was originally explained 
by Barab\'asi and Albert in terms of a growing 
network model\cite{BA}. Later Goh, Kahng and Kim showed 
that a static network model with the scale free 
property could also be constructed\cite{GKK}. 
\par
The connection patterns of the nodes in a complex 
network are represented by the adjacency matrix. 
The spectral (eigenvalue) density of adjacency 
matrices is used to measure the statistical 
features of large complex networks. For example, 
as explained in the following, a power law 
behaviour can be observed. Let us denote 
by $N$ the size of the adjacency matrix. In the 
limit $N \rightarrow \infty$, we obtain an asymptotic 
form of the spectral density. When the mean value $p$ 
of the degree becomes large, Erd\"os and R\'enyi's random 
graphs tend to have the asymptotic spectral density with the 
semicircle shape, as expected from a classical result of 
random matrix theory\cite{WIG}. On the other hand, the tail 
of the asymptotic spectral density for scale free 
networks has a power law behaviour. This power law 
behaviour was analytically confirmed by Dorogovtsev 
et al. for tree-like scale free graphs\cite{DGMS1,DGMS2}. 
Rodgers et al. also found a similar power law for Goh, Kahng 
and Kim's static model\cite{RAKK}. 
\par
Rodgers et al. analysed the spectral density of the static 
model in the limit $p \rightarrow \infty$\cite{RAKK,KK}. 
In their analysis, the replica method in statistical physics was 
employed. A similar usage of the replica method can be traced back 
to Rodgers and Bray's study of sparse random 
matrices\cite{RB}. Recently Semerjian and Cugliandolo\cite{SC} 
proposed a scheme called the effective medium 
approximation (EMA) and elegantly rederived the 
results in \cite{RB}. This scheme was also applied 
to sparse sample covariance matrices 
and analogous results were obtained\cite{NT}. 
\par 
In this paper, we extend the works in \cite{RAKK} 
and \cite{SC} by treating complex networks with a 
finite mean degree $p$. For that purpose, we make 
use of the replica method and EMA for the analysis 
of Goh, Kahng and Kim's static model. Consequently, 
in the asymptotic limit $N \rightarrow \infty$, we 
obtain a new integral equation which determines the 
asymptotic spectral density with a finite $p$. 
We show that known results originally 
found in \cite{RAKK} are rederived in the 
limit $p \rightarrow \infty$. Moreover, using 
the new integral equation, we evaluate the 
$1/p$ corrections to the known results.
\par
This paper is organised as follows. In \S 2, the 
replica method is introduced for Goh, Kahng and Kim's 
static model of scale free networks. In \S 3, using 
EMA, we derive a nonlinear integral equation 
which determines the asymptotic spectral density 
with a mean degree $p$. In the limit $p \rightarrow 
\infty$, the results in \cite{RAKK} are rederived. 
In \S 4, a perturbative method is developed 
to calculate the $1/p$ corrections. In \S 5 and 
\S 6, perturbation terms are analytically 
evaluated around the band center and in the 
tail region.     

\section{Replica Method}
\setcounter{equation}{0}
\renewcommand{\theequation}{2.\arabic{equation}}

Goh, Kahng and Kim's static model of complex networks 
is constructed by the following procedure\cite{GKK}. 
Let us suppose that there are $N$ nodes and consider 
the asymptotic limit $N \rightarrow \infty$. 
Each node $j$ is assigned a probability $P_j$, 
which is normalised as $\sum_{j=1}^N P_j = 1$. 
In order to realize a power-law degree distribution 
with an exponent $\lambda = (1 + \alpha)/\alpha$, we assume that 
\begin{equation}
\label{pj}
P_j = \frac{j^{-\alpha}}{\sum_{j=1}^N j^{-\alpha}} \sim 
(1 - \alpha) N^{\alpha-1} j^{-\alpha}, \ \ \ 0 < \alpha < 1.
\end{equation}
In one step of the procedure, nodes $j$ and $l$ 
are chosen with probabilities $P_j$ and $P_l$, 
respectively. Then the nodes $j$ and $l$ are 
connected by an edge, if $j$ does not coincide 
with $l$ and the nodes are not already connected. 
Repeating such a step $p N/2$ times, one obtains a network 
with a mean degree $p$. The probability that a pair 
of nodes $j$ and $l$ are connected by an edge is   
\begin{equation}
\label{fjl}
f_{jl} = 1 - (1 - 2 P_j P_l)^{p N/2} \sim 
1 - {\rm e}^{- p N P_j P_l}. 
\end{equation}
\par
In this paper, we are interested in the spectral 
density of the adjacency matrix of the network.  
The adjacency matrix $J$ is symmetric ($J_{jl} = 
J_{lj}$) and the elements are independently 
distributed. The p.d.f.(probability distribution 
function) of each element is given by 
\begin{equation}
{\cal P}_{jl}(J_{jl}) = (1 - f_{jl}) \delta(J_{jl}) + f_{jl}  
\delta(J_{jl}-1), \ \ \ j < l.
\end{equation}
Denoting the average over the above probability 
distribution by brackets $\langle \cdot \rangle$, 
we define the spectral density of $J$ as
\begin{equation}
\rho(\mu) = \left\langle \frac{1}{N} \sum_{j=1}^N \delta(\mu - \mu_j) 
\right\rangle,
\end{equation}
where $\mu_1,\mu_2,\cdots,\mu_N$ are the eigenvalues 
of $J$ and $\delta(x)$ is Dirac's delta function. 
\par 
In order to analyse the asymptotic behaviour of 
the spectral density in the limit $N \rightarrow 
\infty$, the partition function
\begin{equation}
Z = \int \prod_{j=1}^N {\rm d}\phi_j \ {\rm exp}\left( \frac{i}{2} \mu 
\sum_{j=1}^N \phi_j^2 - \frac{i}{2} \sum_{jl}^N J_{jl} \phi_j \phi_l \right)
\end{equation}
is useful. In terms of $Z$, the spectral density can 
be rewritten as
\begin{eqnarray}
\rho(\mu) & = & \frac{1}{N \pi} {\rm Im}{\rm Tr}\left\langle 
\{ J - (\mu + i \epsilon) I\}^{-1} \right\rangle \nonumber \\ 
& =  & \frac{2}{N \pi} {\rm Im} \frac{\partial}{\partial \mu} 
\langle \ln Z(\mu + i \epsilon) \rangle.
\end{eqnarray}      
Here $I$ is an identity matrix and $\epsilon$ is a positive 
infinitesimal number. Moreover, the relation
\begin{equation}
\lim_{n \rightarrow 0} \frac{\ln \langle Z^n \rangle}{n} 
= \langle \ln Z \rangle
\end{equation}
can be employed to find
\begin{equation}
\rho(\mu) = \lim_{n \rightarrow 0} \frac{2}{N n \pi} {\rm Im} 
\frac{\partial}{\partial \mu} 
\ln \langle \{ Z(\mu) \}^n \rangle.
\end{equation}
\par
Hence we need to evaluate the average of the $n$-th power of $Z$. 
In terms of the replica variables
\begin{equation}
{\vec \phi}_j = (\phi_j^{(1)},\phi_j^{(2)},\cdots,\phi_j^{(n)})
\end{equation}
and the corresponding measures
\begin{equation}  
{\rm d}{\vec \phi}_j = {\rm d}\phi_j^{(1)}{\rm d}\phi_j^{(2)} 
\cdots {\rm d}\phi_j^{(n)},
\end{equation}
it can be written as 
\begin{equation}
\langle Z^n \rangle = \int \prod_{j=1}^N {\rm d}{\vec \phi}_j \ 
{\rm exp}\left( \frac{i}{2} \mu \sum_{j=1}^N {\vec \phi}_j^2 \right)  
\left\langle {\rm exp}\left( - \frac{i}{2} \sum_{jl}^N J_{jl} 
{\vec \phi}_j \cdot {\vec \phi}_l \right) \right\rangle,
\end{equation} 
where the average is evaluated as
\begin{eqnarray} 
\left\langle {\rm exp}\left( - \frac{i}{2} \sum_{jl}^N J_{jl} 
{\vec \phi}_j \cdot {\vec \phi}_l \right) \right\rangle & = &  
\prod_{j<l}^N \left( \int {\rm d}J_{jl} \ 
{\cal P}_{jl}(J_{jl}) {\rm e}^{ 
-i J_{jl} {\vec \phi}_j \cdot {\vec \phi}_l}  \right) 
\nonumber \\ & = & 
\prod_{j<l}^N \left\{ 
1 + f_{jl}
\left( {\rm e}^{-i 
{\vec \phi}_j \cdot {\vec \phi}_l} - 1 \right)
\right\}. 
\end{eqnarray}
Then one obtains an asymptotic estimate in the limit 
$N \rightarrow \infty$ as 
\begin{equation} 
\ln \left\langle {\rm exp}\left( - \frac{i}{2} \sum_{jl}^N J_{jl} 
{\vec \phi}_j \cdot {\vec \phi}_l \right) \right\rangle \sim   
p N \sum_{j<l}^N P_j P_l \left( {\rm e}^{-i 
{\vec \phi}_j \cdot {\vec \phi}_l} - 1 \right).
\end{equation}
It was shown in \cite{KRKK} that the remainder term of 
this asymptotic estimate was $O(1)$ for $0 < \alpha < 1/2$, 
$O((\ln N)^2)$ for $\alpha = 1/2$ and $O(N^{2 - (1/\alpha)} 
\ln N)$ for $1/2 < \alpha < 1$. Moreover we define 
\begin{equation}
{\tilde c}_j({\vec \phi}) = \delta({\vec \phi}- {\vec \phi}_j),
\end{equation}
so that
\begin{eqnarray} 
& & \left\langle {\rm exp}\left( - \frac{i}{2} \sum_{jl}^N J_{jl} 
{\vec \phi}_j \cdot {\vec \phi}_l \right) \right\rangle 
\nonumber \\ & \sim & {\rm exp}\left\{ 
\frac{p N}{2} \sum_{jl}^N P_j P_l \int {\rm d}{\vec \phi} 
\int {\rm d}{\vec \psi} \ 
{\tilde c}_j({\vec \phi}) {\tilde c}_l({\vec \psi}) \left( 
{\rm e}^{- i {\vec \phi} \cdot {\vec \psi}}  
- 1 \right) \right\}.
\end{eqnarray}
Then, introducing auxiliary functions $c_j({\vec \phi})$, 
we find  
\begin{eqnarray}
\langle Z^n \rangle & \sim & 
\int \prod_{j=1}^N {\rm d}{\vec \phi}_j \ 
{\rm exp}\left\{ 
\frac{i}{2} \mu \sum_{j=1}^N \int {\rm d}{\vec \phi} \  
{\tilde c}_j({\vec \phi}) {\vec \phi}^2 \right\} \nonumber \\ 
& \times & {\rm exp}\left\{ 
\frac{p N}{2} \sum_{jl}^N P_j P_l 
\int {\rm d}{\vec \phi} \int {\rm d}{\vec \psi} \  
{\tilde c}_j({\vec \phi}) {\tilde c}_l({\vec \psi}) \left( 
{\rm e}^{- i {\vec \phi} \cdot {\vec \psi}}  
- 1 \right) \right\} 
\nonumber \\ & = & 
\int \prod_{j=1}^N {\rm d}{\vec \phi}_j \int 
\prod_{j=1}^N {\cal D}c_j({\vec \phi}) \prod_{j=1}^N \prod_{\vec \phi} 
\delta(c_j({\vec \phi}) - {\tilde c}_j({\vec \phi})) {\rm e}^{S_1 + S_2},
\end{eqnarray} 
where
\begin{equation}
S_1 = \frac{i}{2} \mu \sum_{j=1}^N \int {\rm d}{\vec \phi} 
\ c_j({\vec \phi}) {\vec \phi}^2 
\end{equation}
and
\begin{equation}
S_2 = \frac{p N}{2} \sum_{jl}^N P_j P_l 
\int {\rm d}{\vec \phi} \int {\rm d}{\vec \psi} 
\ c_j({\vec \phi}) \ c_l({\vec \psi}) \left( 
{\rm e}^{- i {\vec \phi} \cdot {\vec \psi}}  
- 1 \right).
\end{equation}
Here the functional integration is taken over $c_j({\vec \phi})$ 
satisfying
\begin{equation}
\int {\rm d}{\vec \phi} \ c_j({\vec \phi}) = 1.
\end{equation}
Let us note
\begin{eqnarray}
\label{e1} 
& &  \int \prod_{j=1}^N {\rm d}{\vec \phi}_j \prod_{j=1}^N \prod_{\vec \phi} 
\delta(c_j({\vec \phi}) - {\tilde c}_j({\vec \phi})) \nonumber \\ 
& = &  \int \prod_{j=1}^N {\rm d}{\vec \phi}_j 
\int \prod_{j=1}^N {\cal D}a_j({\vec \phi}) \ {\rm exp} \left[
2 \pi i \sum_{j=1}^N \int {\rm d}{\vec \phi} \ a_j({\vec \phi})
\left\{ c_j({\vec \phi})-{\tilde c}_j({\vec \phi}) \right\} \right]
\nonumber \\ 
& = &  \int \prod_{j=1}^N {\cal D}a_j({\vec \phi}) \ {\rm exp}\left[ 
\sum_{j=1}^N \left\{ 2 \pi i \int 
{\rm d}{\vec \phi} \ a_j({\vec \phi}) c_j({\vec \phi}) - F_j \right\} \right], 
\end{eqnarray}
where
\begin{eqnarray}
F_j & = & - \ln \int {\rm d}{\vec \phi}_j \ {\rm exp}\left\{ 
- 2 \pi i \int {\rm d}{\vec \phi} \ a_j({\vec \phi}) 
{\tilde c}_j({\vec \phi}) \right\} \nonumber \\ 
& = &  - \ln \int {\rm d}{\vec \phi}_j \ {\rm exp}\left\{ 
- 2 \pi i a_j({\vec \phi}_j) \right\}. 
\end{eqnarray}
In the limit $N \rightarrow \infty$, the functional integral (\ref{e1}) 
is dominated by the contribution from the neighbourhood of the 
extremum satisfying
\begin{equation}
\frac{\delta}{\delta a_j({\vec \phi})} \left\{  
2 \pi i \int {\rm d}{\vec \phi} \ a_j({\vec \phi}) c_j({\vec \phi}) 
- F_j \right\} = 2 \pi i c_j({\vec \phi}) - 2 \pi i {\rm e}^{ 
- 2 \pi i a_j({\vec \phi}) + F_j } = 0,
\end{equation} 
so that
\begin{equation}
- \int {\rm d}{\vec \phi} \ c_j({\vec \phi}) \ln c_j({\vec \phi}) = 2 \pi i
\int {\rm d}{\vec \phi} \ a_j({\vec \phi}) c_j({\vec \phi}) - F_j.
\end{equation}
We thus obtain
\begin{equation} 
\int \prod_{j=1}^N {\rm d}{\vec \phi}_j \prod_{j=1}^N \prod_{\vec \phi} 
\delta(c_j({\vec \phi}) - {\tilde c}_j({\vec \phi})) \nonumber \\ 
 \sim {\rm exp}\left\{ -  
\sum_{j=1}^N \int {\rm d}{\vec \phi} \ c_j({\vec \phi}) \ln c_j({\vec \phi}) \right\},  
\end{equation}
from which it follows that 
\begin{equation}
\label{e2}
\langle Z^n \rangle \sim \int \prod_{j=1}^N 
{\cal D}c_j({\vec \phi}) \ {\rm e}^{S_0 + S_1 + S_2} 
\end{equation} 
with
\begin{equation}
S_0 = - \sum_{j=1}^N \int {\rm d}{\vec \phi} 
\ c_j({\vec \phi}) \ln c_j({\vec \phi}).
\end{equation}

\section{Integral Equation}
\setcounter{equation}{0}
\renewcommand{\theequation}{3.\arabic{equation}}

In the limit $N \rightarrow \infty$, the dominant contribution to the 
functional integral (\ref{e2}) again comes from the neighbourhood of 
the extremum. In order to work out the extremum, we 
replace $c_j({\vec \phi})$ with a Gaussian ansatz 
\begin{equation}
c_j({\vec \phi}) = \frac{1}{(2 \pi i \sigma_j)^{n/2}} {\rm exp}
\left(  - \frac{{\vec \phi}^2}{2 i \sigma_j} \right). 
\end{equation}
Then the extremum condition can be written as
\begin{equation}
\label{e3}
\frac{\partial}{\partial \sigma_k} (S_0 + S_1 + S_2) = 0.
\end{equation}
This variational scheme based on the Gaussian ansatz 
is called the effective medium approximation (EMA)\cite{SC}.
\par
Using the identity 
\begin{equation}
\int {\rm d}{\vec \phi} \ {\vec \phi}^2 c_j({\vec \phi}) = n i \sigma_j,
\end{equation}
we can readily find
\begin{equation}
S_0 = \frac{n}{2} \sum_{j=1}^N \ln(2 \pi i \sigma_j) + \frac{Nn}{2}, 
\ \ \ S_1 = - \frac{\mu n}{2} \sum_{j=1}^N \sigma_j,
\end{equation}
so that
\begin{equation}
\frac{\partial S_0}{\partial \sigma_k} = \frac{n}{2 \sigma_k}, \ \ \ 
\frac{\partial S_1}{\partial \sigma_k} = - \frac{\mu n}{2}. 
\end{equation} 
It is straightforward to obtain 
\begin{eqnarray}
S_2 & = & -\frac{N p}{2} \left\{  1  -  \sum_{jl}^N P_j P_l 
\int {\rm d}{\vec \phi} {\rm d}{\vec \psi} 
\frac{{\rm exp}\left( -\frac{{\vec \phi}^2}{2 i \sigma_j} - 
\frac{{\vec \psi}^2}{2 i \sigma_l} 
- i {\vec \phi} \cdot {\vec \psi} 
\right) }{(2 
\pi i \sigma_j)^{n/2}(2 \pi i \sigma_l)^{n/2}} 
\right\} \nonumber \\ 
& = & -\frac{N p}{2} \left[ 1  -  \sum_{jl}^N P_j P_l 
\int {\rm d}{\vec \psi} \frac{1}{(2 \pi i \sigma_l)^{n/2}} 
{\rm exp}\left\{ - \frac{1 - \sigma_j \sigma_l}{2 i \sigma_l} {\vec \psi}^2 
\right\} \right] \nonumber \\ 
& = & -\frac{Np}{2} \left\{ 1 - \sum_{jl}^N \frac{P_j P_l}{(1 - \sigma_j 
\sigma_l)^{n/2}} \right\}. 
\end{eqnarray} 
Therefore
\begin{equation}
\frac{\partial S_2}{\partial \sigma_k} = \frac{Npn}{2} P_k 
\sum_{j=1}^N \frac{P_j \sigma_j}{(1 - \sigma_j \sigma_k)^{(n/2)+1}}. 
\end{equation} 
Then the extremum condition (\ref{e3}) yields
\begin{equation}
\label{e4}
\mu - \frac{1}{\sigma_k} - N p P_k \sum_{j=1}^N \frac{P_j \sigma_j}{
1 - \sigma_j \sigma_k} = 0
\end{equation}
in the limit $n \rightarrow 0$. Putting (\ref{pj}) into (\ref{e4}), 
we find   
\begin{equation}
\label{sumeq}
0 = \mu - \frac{1}{\sigma_k} - p (1 - \alpha)^2 \left( \frac{k}{N} 
\right)^{-\alpha} \frac{1}{N} \sum_{j=1}^N \frac{(j/N)^{-\alpha} 
\sigma_j}{ 1 - \sigma_j \sigma_k}. 
\end{equation}
In the limit $N \rightarrow \infty$, it follows that
\begin{equation}
\label{ieq}
\sigma(x) x^{-\alpha} = \frac{1}{\displaystyle 
\mu x^{\alpha} - p (1 - \alpha)^2 
\int_0^1 \frac{y^{-\alpha} 
\sigma(y)}{ 1 - \sigma(x) \sigma(y)} {\rm d}y}
\end{equation}
with $x = k/N$. The function $\sigma(x)$ is determined 
by this integral equation. 
\par
The spectral density can be evaluated as
\begin{eqnarray}
\label{rhomu}
\rho(\mu) & = & \lim_{n \rightarrow 0} \frac{2}{N n \pi} {\rm Im} 
\frac{\partial}{\partial \mu} 
\ln \langle \{ Z(\mu) \}^n \rangle \nonumber \\ 
 & \sim  & \lim_{n \rightarrow 0} \frac{2}{N n \pi} {\rm Im} 
\frac{\partial}{\partial \mu} (S_0 + S_1 + S_2) \nonumber \\   
 & = & - \frac{1}{N \pi} {\rm Im} \sum_{j=1}^N \sigma_j,  
\end{eqnarray}
so that
\begin{equation}
\rho(\mu) = - \frac{1}{\pi} {\rm Im} \int_0^1 \sigma(x) {\rm d}x
\end{equation}
in the limit $N \rightarrow \infty$. The asymptotic spectral 
density is thus written in terms of the solution of 
the integral equation (\ref{ieq}).
\par
Using (\ref{pj}) and (\ref{fjl}), we generated adjacency 
matrices $J$ with $N = 1000$ and evaluated the spectral density 
(NUMERICAL) by numerical diagonalization. On the other hand, the 
EMA spectral density (EMA) was calculated by solving (\ref{sumeq}) 
with $N = 1000$ by numerical iteration and by putting the solution 
into (\ref{rhomu}). We show the both results for $p = 10$ and 
$\alpha = 0.5$ in Figure 1. The agreement seems good enough to 
demonstrate the validity of EMA.              

\begin{figure}[h]
\epsfxsize=14cm
\centerline{\epsfbox{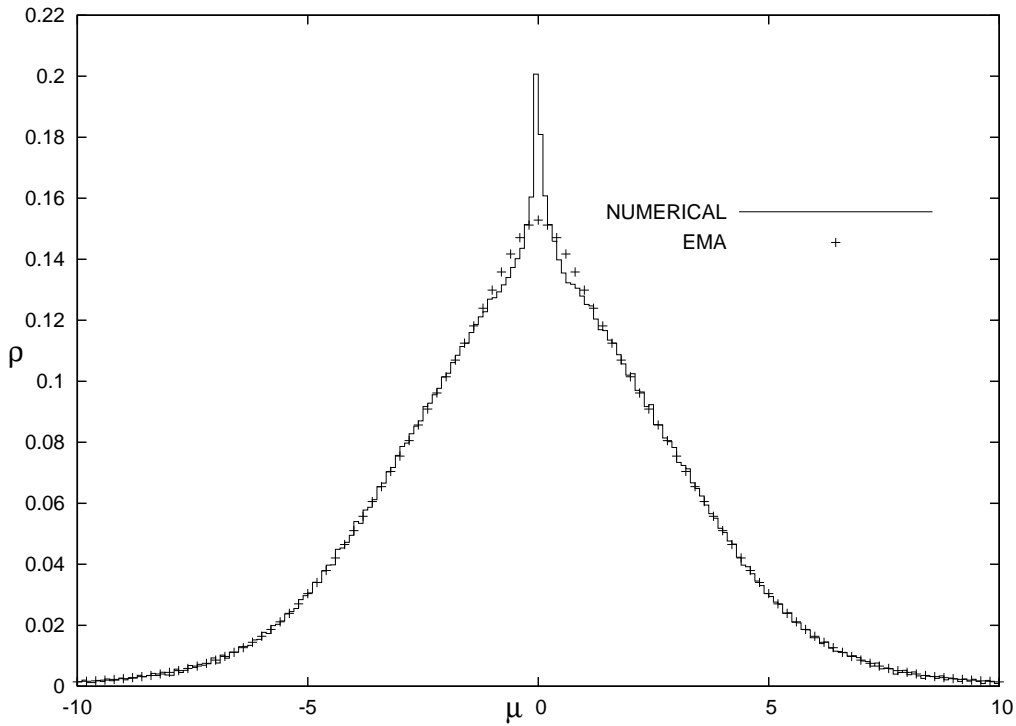}}
\caption{Comparison of a numerically generated 
spectral density (average over $100$ samples, $N = 1000$) 
and an EMA solution ($p = 10, \alpha = 0.5$).}
\end{figure}
 
\section{Perturbation Theory}
\setcounter{equation}{0}
\renewcommand{\theequation}{4.\arabic{equation}}

In order to analytically evaluate the EMA 
spectral density, we start from the integral 
equation (\ref{ieq}) and develop a perturbation 
theory for large $p$. Introducing scalings 
\begin{equation}
\sigma(x) \rightarrow \frac{1}{\sqrt{p}} s(x), \ \ \ 
\mu \rightarrow \sqrt{p} m,
\end{equation} 
we obtain
\begin{equation}
\label{sieq}
s(x) x^{-\alpha} = \frac{1}{\displaystyle 
m x^{\alpha} - (1 - \alpha)^2 
\int_0^1 \frac{y^{-\alpha} 
s(y)}{\displaystyle 1 - \frac{1}{p} s(x) s(y)} {\rm d}y}.
\end{equation}
Let us expand $s(x)$ in terms of $1/p$ as
\begin{equation}
s(x) = s_0(x) + \frac{1}{p} s_1(x) + O\left(\frac{1}{p^2}\right),
\end{equation}
so that the unperturbed term $s_0(x)$ satisfies
\begin{equation}
\label{s0x}
s_0(x) x^{-\alpha} = \frac{1}{\displaystyle 
m x^{\alpha} - (1 - \alpha)^2 
\int_0^1 y^{-\alpha} s_0(y){\rm d}y}.
\end{equation}
This integral equation for the unperturbed term was originally 
found in \cite{RAKK}.  
\par
Then, expanding the RHS of (\ref{sieq}) in terms of $1/p$, 
we find the relation
\begin{eqnarray}
& & \left\{ s_0(x) + \frac{1}{p} s_1(x) \right\} x^{-\alpha} \nonumber \\ 
& \sim & s_0(x) x^{-\alpha} + 
\frac{(1 - \alpha)^2}{p} \{s_0(x)\}^2 x^{- 2 \alpha} \int_0^1  
\left[ s_1(y) + s_0(x) \{s_0(y)\}^2 \right] y^{-\alpha} {\rm d}y 
\nonumber \\ 
\end{eqnarray}
in the limit $p \rightarrow \infty$. Therefore, 
the perturbation term $s_1(x)$ can be evaluated as
\begin{eqnarray}
\label{s1}      
s_1(x) x^{-\alpha} & = &  
(1 - \alpha)^2 \{s_0(x)\}^2 x^{- 2 \alpha} \int_0^1 
s_1(y) y^{-\alpha} {\rm d}y 
\nonumber \\ & + &  
(1 - \alpha)^2 \{s_0(x)\}^3 x^{- 2 \alpha} \int_0^1 
\{s_0(y)\}^2 y^{-\alpha} {\rm d}y,  
\end{eqnarray}
so that
\begin{equation}
\label{s1int1}
\int_0^1 s_1(x) {\rm d}x = (1 - \alpha)^2 I_1 (D + J_1).
\end{equation}
Here 
\begin{equation}
I_1 = \int_0^1 \{s_0(x)\}^2 x^{-\alpha} {\rm d}x,
\ \ \ J_1 = \int_0^1 \{s_0(x)\}^3 x^{-\alpha} {\rm d}x
\end{equation}
and
\begin{equation}
D = \int_0^1 s_1(x) x^{-\alpha} {\rm d}x.
\end{equation}
Integrating the both sides of (\ref{s1}) over $x$ 
from $0$ to $1$, we find
\begin{eqnarray}
D & = & (1 - \alpha)^2 D \int_0^1 \{s_0(x)\}^2 x^{- 2 \alpha} 
{\rm d}x + (1 - \alpha)^2 I_1 \int_0^1 \{s_0(x)\}^3 x^{- 2 \alpha} 
{\rm d}x \nonumber \\ 
& = & (1 - \alpha)^2 D I_2 + (1 - \alpha)^2 I_1 J_2, 
\end{eqnarray}
where
\begin{equation}
I_2 = \int_0^1 \{s_0(x)\}^2 x^{- 2 \alpha} {\rm d}x,
\ \ \ J_2 = \int_0^1 \{s_0(x)\}^3 x^{-2 \alpha} {\rm d}x.
\end{equation}
Therefore $D$ can be evaluated as 
\begin{equation}
\label{dintegral}
D = \frac{(1 - \alpha)^2 I_1 J_2}{1 - 
(1 - \alpha)^2 I_2}, 
\end{equation} 
so that
\begin{equation}
\label{s1int2}
\int_0^1 s_1(x) {\rm d}x = (1 - \alpha)^2 I_1 \left\{ 
\frac{(1 - \alpha)^2 I_1 J_2}{1 - 
(1 - \alpha)^2 I_2} + J_1 \right\}. 
\end{equation}
\par
Let us summarise the scheme of the perturbation. 
First we evaluate $s_0(x)$ by solving the 
integral equation (\ref{s0x}). Next we calculate 
\begin{equation}
\label{imints0}
{\rm Im} \int_0^1 s_0(x) {\rm d}x 
\end{equation}
and the other integrals $I_1$, $J_1$, $I_2$ and $J_2$.
Then, using the relation (\ref{s1int2}), we are able to 
compute 
\begin{equation}
\label{imints1}
{\rm Im} \int_0^1 s_1(x) {\rm d}x. 
\end{equation}
In terms of the integrals (\ref{imints0}) 
and (\ref{imints1}), the asymptotic spectral 
density is expressed as
\begin{eqnarray}
\label{rhoexpand}
\rho(\mu) & \sim & - \frac{1}{\pi \sqrt{p}} 
\left\{{\rm Im} \int_0^1 s_0(x) {\rm d}x 
+ \frac{1}{p} {\rm Im} \int_0^1 s_1(x) {\rm d}x 
+ O\left(\frac{1}{p^2}\right) \right\}
\end{eqnarray}
in the limit $N \rightarrow \infty$.
\par
In the following sections, we evaluate 
the small and large $m$ expansions of 
the integrals (\ref{imints0}) and 
(\ref{imints1}), so that (\ref{rhoexpand}) 
gives the asymptotic behaviour of the 
EMA spectral density around the band 
center and in the tail region.   

\section{Band Center}
\setcounter{equation}{0}
\renewcommand{\theequation}{5.\arabic{equation}}

Let us evaluate the EMA spectral density around 
the band center, by calculating the small $m$ expansions 
of the integrals (\ref{imints0}) and (\ref{imints1}). 
Noting the dependence of $s_0(x)$ on $m$, we expand it as 
\begin{equation}
s_0(x) = g_0(x) + m g_1(x) + m^2 g_2(x) + O\left( m^3 \right).
\end{equation}
\par
In order to compute the central value 
$g_0(x)$, we put $m=0$ in (\ref{s0x}) 
and find  
\begin{equation}
g_0(x) x^{-\alpha} = - \frac{1}{\displaystyle (1 - \alpha)^2 
\int_0^1 y^{-\alpha} g_0(y){\rm d}y}.
\end{equation}
Integrating the both sides in $x$ from $0$ to $1$, 
we obtain 
\begin{equation}
\int_0^1 y^{-\alpha} g_0(y) {\rm d}y = - i \frac{1}{1 - \alpha},
\end{equation}
so that
\begin{equation}
g_0(x) = - i \frac{1}{1 - \alpha} x^{\alpha}.
\end{equation}
\par
Let us next evaluate $g_1(x)$ and $g_2(x)$. 
We expand (\ref{s0x}) in terms of $m$ and find 
\begin{eqnarray}
\label{g1g2}
& & \left\{ g_0(x) + m g_1(x) + m^2 g_2(x) \right\} x^{-\alpha} 
\nonumber \\ & \sim & 
g_0(x) x^{-\alpha} \left[ 1 - m g_0(x) x^{-\alpha} \left\{ 
x^{\alpha} - (1 - \alpha)^2 \int_0^1 g_1(y) y^{-\alpha} {\rm d}y \right\} 
\right. \nonumber \\ & + & \left. 
m^2 g_0(x) x^{-\alpha} (1 - \alpha)^2 \int_0^1 g_2(y) y^{-\alpha} {\rm d}y 
\right. \nonumber \\ & + & \left. m^2 \{ g_0(x) \}^2 x^{- 2 \alpha} \left\{ 
x^{\alpha} - (1 - \alpha)^2 \int_0^1 g_1(y) y^{-\alpha} {\rm d}y \right\}^2 
\right]    
\end{eqnarray}
in the limit $m \rightarrow 0$. The linear term in $m$ reads
\begin{eqnarray}
g_1(x) x^{-\alpha} 
& = &  - \{ g_0(x) \}^2 x^{-2 \alpha} \left\{ 
x^{\alpha} - (1 - \alpha)^2 \int_0^1 g_1(y) y^{-\alpha} {\rm d}y \right\} 
\nonumber \\ 
& = &  \frac{x^{\alpha}}{(1 - \alpha)^2} -  
\int_0^1 g_1(y) y^{-\alpha} {\rm d}y. 
\end{eqnarray}
We again integrate the both sides in $x$ from $0$ to $1$ and find 
\begin{equation}
\int_0^1 g_1(y) y^{-\alpha} {\rm d}y = \frac{1}{2 (1 - \alpha)^2 
( 1 + \alpha)},  
\end{equation}
so that
\begin{equation}
g_1(x) x^{-\alpha} = \frac{x^{\alpha}}{(1 - \alpha)^2} -  
\frac{1}{2 (1 - \alpha)^2 (1 + \alpha)}.
\end{equation}
The quadratic term in $m$ is similarly extracted from 
(\ref{g1g2}) as
\begin{eqnarray}
g_2(x) x^{-\alpha} 
& = &  
\{ g_0(x) \}^2 x^{-2 \alpha} 
(1 - \alpha)^2 \int_0^1 g_2(y) y^{-\alpha} {\rm d}y 
 \nonumber \\ & + & \{ g_0(x) \}^3 x^{- 3 \alpha} \left\{ 
x^{\alpha} - (1 - \alpha)^2 \int_0^1 g_1(y) y^{-\alpha} {\rm d}y \right\}^2  
\nonumber \\ 
& =  & - \int_0^1 g_2(y) y^{-\alpha} {\rm d}y + i \frac{1}{(1 - \alpha)^3} 
\left\{ x^{\alpha} - \frac{1}{2 (1 + \alpha)} \right\}^2.
\end{eqnarray}
Integrating the both sides in $x$ from $0$ to $1$ yields  
\begin{equation}
\int_0^1 g_2(y) y^{-\alpha} {\rm d}y = i \frac{1}{2 (1 - \alpha)^3} 
\left\{ \frac{1}{1 + 2 \alpha}- \frac{3}{4} \frac{1}{(1 + \alpha)^2} 
\right\},  
\end{equation}
from which it follows that
\begin{equation}
g_2(x) x^{-\alpha} = i \frac{1}{2 (1 - \alpha)^3} \left\{ 
2 x^{2 \alpha} - \frac{2 x^{\alpha}}{1 + \alpha} + \frac{5}{4} 
\frac{1}{(1 + \alpha)^2} - \frac{1}{1 + 2 \alpha} \right\}.
\end{equation}
Thus we obtain the small $m$ expansion of $s_0(x)$:
\begin{eqnarray}
& & s_0(x) x^{-\alpha} = - i \frac{1}{1 - \alpha} + \frac{m}{(1 - \alpha)^2} 
\left\{ x^{\alpha} - \frac{1}{2 (1 + \alpha)} \right\} 
\nonumber \\ & + & i \frac{m^2}{2 (1 - \alpha)^3} \left\{ 
2 x^{2 \alpha} - \frac{2 x^{\alpha}}{1 + \alpha} 
+ \frac{5}{4} \frac{1}{(1 + \alpha)^2} - \frac{1}{1 + 2 \alpha} \right\} 
+ O\left( m^3 \right), \nonumber \\ 
\end{eqnarray}
so that
\begin{equation}
\label{ints0x}
{\rm Im} \int_0^1 s_0(x) {\rm d}x = - \frac{1}{1 - \alpha^2} 
+ \frac{m^2(1 + 5 \alpha + 
18 \alpha^2 + 20 \alpha^3 + 16 \alpha^4)}{8 (1 - \alpha^2)^3 (1 + 2 \alpha)(1 + 3 \alpha)} + O\left( 
m^4 \right),
\end{equation}
which yields the spectral density in the limit $p \rightarrow \infty$. 
This result for the unperturbed term was first derived in \cite{RAKK}. 
\par
We are now in a position to evaluate the perturbation term. The 
integral $D$ can be computed from (\ref{dintegral}) as
\begin{eqnarray}
D & = & -i \frac{1}{2 (1 - \alpha)^3 (1 + \alpha)^2} + 
\frac{m (2 + 4 \alpha + 5 \alpha^2)}{2 (1 - \alpha)^4 (1 + \alpha)^3 (1 + 2 \alpha)} 
\nonumber \\ 
& + & 3 i \frac{m^2 (5 + 35 \alpha + 128 \alpha^2 + 252 \alpha^3 + 256 \alpha^4 + 144 \alpha^5)}{
16 (1 - \alpha)^5 (1 + \alpha)^4 (1 + 2 \alpha)^2 (1 + 3 \alpha)} + O\left( m^3 \right). 
\nonumber \\ 
\end{eqnarray} 
Then it follows from (\ref{s1int1}) that
\begin{eqnarray}
\label{ints1x}
& & {\rm Im} \int_0^1 s_1(x) {\rm d}x = 
- \frac{1 + 2 \alpha + 2 \alpha^2}{2 (1 + 2 \alpha) (1 - \alpha^2)^3} 
\nonumber \\ 
& + & 
\frac{3 m^2 (5 + 55 \alpha + 314 \alpha^2 + 1034 
\alpha^3 + 2068 \alpha^4 + 2648 \alpha^5 + 
1920 \alpha^6 + 768 \alpha^7)}{16 (1 + 2 \alpha)^2 (1 - \alpha^2)^5 (1 + 7 \alpha + 12 \alpha^2)} 
\nonumber \\  & + & O\left( m^4 \right).
\end{eqnarray}
\par
As mentioned in the Introduction, Rodgers and Bray\cite{RB} 
analysed sparse random matrices and derived a formula for the 
asymptotic spectral density 
\begin{equation}
\label{rb}
\rho(\mu) \sim \frac{2}{\pi (\mu_c)^2} \{(\mu_c)^2 - \mu^2\}^{1/2} 
\left[ 1 + \frac{1}{p} \left\{ 1 - 4 \frac{\mu^2}{(\mu_c)^2} \right\} 
 + O\left( \frac{1}{p^2} \right) \right]   
\end{equation}
with $\mu_c^2 = 4 \{ p + 1 + O(1/p) \}$. Sparse random matrices 
in \cite{RB} give the adjacency matrices of Erd\"os and R\'enyi's 
random graphs, which correspond to the limit $\alpha \rightarrow 0$ 
of Goh, Kahng and Kim's complex networks. From (\ref{ints0x}) 
and (\ref{ints1x}), we obtain 
\begin{equation}
{\rm Im} \int_0^1 s_0(x) {\rm d}x = -1 + \frac{1}{8} m^2  + O\left(m^4 \right)
\end{equation}
and
\begin{equation}
{\rm Im} \int_0^1 s_1(x) {\rm d}x 
= - \frac{1}{2} + \frac{15}{16} m^2 + O\left( m^4 \right)
\end{equation}
in the limit $\alpha \rightarrow 0$. It follows that 
\begin{equation}
\rho(\mu) \sim \frac{1}{\pi \sqrt{p}} \left[ 1 - 
\frac{m^2}{8} + O\left(m^4 \right) 
+ \frac{1}{p} \left\{  
\frac{1}{2} - \frac{15}{16} m^2 + O\left( m^4 \right)
\right\}
+ O\left(\frac{1}{p^2}\right) \right] 
\end{equation}
in agreement with (\ref{rb}), as expected. 
 
\section{Tail Region}
\setcounter{equation}{0}
\renewcommand{\theequation}{6.\arabic{equation}}

In order to analyse the EMA spectral density in the 
tail region, we need to work out the asymptotic formulae 
in the limit $m \rightarrow \infty$ for the integrals 
(\ref{imints0}) and (\ref{imints1}). 
\par
To begin with, we define $c = a - i b$ (with real $a$ 
and $b$) as
\begin{equation}
c = (1 - \alpha)^2 \int_0^1 s_0(x) x^{-\alpha} {\rm d}x
\end{equation}
and rewrite (\ref{s0x}) as 
\begin{equation}
s_0(x) x^{-\alpha} = \frac{1}{m x^{\alpha} - c}.
\end{equation}
Integrating over $x$ from $0$ to $1$, one obtains 
\begin{equation}  
\frac{c}{(1 - \alpha)^2} = \int_0^1 \frac{1}{m x^{\alpha} - c} 
{\rm d}x. 
\end{equation}
This integral equation determines the parameter $c$. 
In \cite{RAKK}, asymptotic formulae 
\begin{eqnarray}
\label{ab}
a & \sim & \frac{1 - \alpha}{m}, \nonumber \\ 
b & \sim & \pi \frac{(1 - \alpha)^{(1 + \alpha)/\alpha}}{\alpha} 
\frac{1}{m^{(2-\alpha)/\alpha}}
\end{eqnarray}
were derived in the limit $m \rightarrow \infty$. It follows that
\begin{eqnarray}
{\rm Im} \int_0^1 s_0(x) {\rm d}x 
& = & {\rm Im} \int_0^1 \frac{x^{\alpha}}{m x^{\alpha} - c} {\rm d}x
\nonumber \\ & \sim & 
- 2 \pi \frac{(1 - \alpha)^{1/\alpha}}{\alpha} 
\frac{1}{m^{(2+\alpha)/\alpha}},
\end{eqnarray}
which gives the tail behaviour of the spectral density in 
the limit $p \rightarrow \infty$. 
\par
Let us evaluate the $1/p$ correction to this asymptotic 
formula. Using the expressions
\begin{equation}
I_1 = \int_0^1 \frac{x^{\alpha}}{(m x^{\alpha} - c)^2} 
{\rm d}x, \ \ \   I_2 = \int_0^1 \frac{1}{(m x^{\alpha} - c)^2} 
{\rm d}x, 
\end{equation}
\begin{equation}
J_1 = \int_0^1 \frac{x^{2 \alpha}}{(m x^{\alpha} - c)^3} 
{\rm d}x, \ \ \   
J_2 = \int_0^1 \frac{x^{\alpha}}{(m x^{\alpha} - c)^3} 
{\rm d}x
\end{equation}  
and (\ref{ab}), we can derive the asymptotic formulae 
\begin{eqnarray}
{\rm Re}\ I_1 & \sim & \frac{1}{1 - \alpha} \frac{1}{m^2}, \nonumber \\ 
{\rm Im}\ I_1 & \sim & - \pi 
\frac{(1 - \alpha)^{(1 - \alpha)/\alpha}}{\alpha^2} \frac{1}{
m^{2/\alpha}},
\end{eqnarray}  
\begin{eqnarray}
{\rm Re}\ J_1 & \sim & \frac{1}{1 - \alpha} \frac{1}{m^3}, \nonumber \\ 
{\rm Im}\ J_1 & \sim & - \frac{\pi}{2}  
\frac{(1 + \alpha) \ (1 - \alpha)^{(1 - \alpha)/\alpha}}{\alpha^3} 
\frac{1}{m^{(2+\alpha)/\alpha}},
\end{eqnarray}  
\begin{eqnarray}
{\rm Im}\ I_2 \sim - \pi  
\frac{(1 - \alpha)^{(1 - \alpha)/\alpha}}{\alpha^2} 
\frac{1}{m^{2(1 -\alpha)/\alpha}}
\end{eqnarray}
and  
\begin{eqnarray}
{\rm Im}\ J_2 \sim - \frac{\pi}{2}   
\frac{(1 - \alpha)^{(1 - \alpha)/\alpha}}{\alpha^3} 
\frac{1}{m^{(2 -\alpha)/\alpha}}
\end{eqnarray}
in the limit $m \rightarrow \infty$. Moreover, 
omitting logarithmic factors of the form 
$\ln m$, we find the asymptotic estimates
\begin{equation}
{\rm Re}\ I_2 \sim \left\{ \begin{array}{ll} 
O(m^{-2}), & \alpha \leq 1/2, \\ 
O(m^{2(\alpha - 1)/\alpha}), & \alpha > 1/2 
\end{array} \right.
\end{equation}
and     
\begin{equation}
{\rm Re}\ J_2 \sim \left\{ \begin{array}{ll} 
O(m^{-3}), & \alpha \leq 1/2, \\ 
O(m^{(\alpha - 2)/\alpha}), & \alpha > 1/2. 
\end{array} \right.
\end{equation}     
\par
Putting the above asymptotic formulae into 
(\ref{s1int2}), we obtain
\begin{equation}
{\rm Im} \int_0^1 s_1(x) {\rm d}x \sim 
- \pi \frac{(1 + \alpha) \ (1 - \alpha)^{1/\alpha}}{\alpha^3} 
\frac{1}{m^{(2 + 3 \alpha)/\alpha}}.
\end{equation} 

\section{Discussion}

Using the replica method and effective medium 
approximation (EMA) in statistical physics, 
we evaluated the spectral density for the 
adjacency matrix of Goh, Kahng and Kim's 
model of complex networks with a finite 
mean degree $p$. The EMA result 
was compared with numerically generated 
spectral density. In the limit $p \rightarrow 
\infty$, known results derived in \cite{RAKK} 
were reproduced. Moreover, perturbative 
analytic formulae were presented for the 
EMA solution in the forms of $1/p$ expansions. As shown in 
Figure 1, although the agreement of the EMA spectral density 
with numerical data was fairly good, a significant 
discrepancy was observed around the band center. 
In order to improve the agreement, it seems necessary 
to develop more accurate schemes, such as a non-perturbative 
technique\cite{RB} or single defect approximation\cite{SC,NT,BM}. 
The discrepancy around the band center might be related to the 
occurrence of the eigenvector localisation. In connection to 
the localisation, we expect that further studies on the local 
properties of the spectra, such as the distribution of the 
eigenvalue spacings\cite{BS} or the largest eigenvalues\cite{CLV}, 
will be illuminating.       

\section*{Acknowledgements}

One of the authors (T.N.) thanks Prof. Toshiyuki Tanaka for 
valuable discussions. This work was partially supported by 
the Grant-in-Aid for Scientific Research, MEXT, Japan 
(Nos.~16740224).


\begin{thebibliography}{notitle}

\bibitem{ER} P. Erd\"os and A. R\'enyi, Publ. Math. Inst. Hung. 
Acad. Sci. Ser. A {\bf 5} (1960) 17.
\bibitem{WS} D.J. Watts and S.H. Strogatz, Nature {\bf 393} 
(1998) 440.
\bibitem{BA} A.-L. Barab\'asi and R. Albert, Science {\bf 286} 
(1999) 509.
\bibitem{GKK}
K.-I. Goh, B. Kahng and D. Kim, Phys. Rev. Lett. {\bf 87} (2001) 278701.
\bibitem{WIG}
E.P. Wigner, Ann. Math. {\bf 67} (1958) 325.  
\bibitem{DGMS1}
S.N. Dorogovtsev, A.V. Goltsev, J.F.F. Mendes and A.N. Samukhin, 
Phys. Rev. {\bf E68} (2003) 046109.
\bibitem{DGMS2}
S.N. Dorogovtsev, A.V. Goltsev, J.F.F. Mendes and A.N. Samukhin, 
Physica {\bf A338} (2004) 76.
\bibitem{RAKK}
G.J. Rodgers, K. Austin, B. Kahng and D. Kim, J. Phys. {\bf A38} 
(2005) 9431.
\bibitem{KK}
D. Kim and B. Kahng, Chaos, {\bf 17} (2007) 026115. 
\bibitem{RB}
G.J. Rodgers and A.J. Bray, Phys. Rev. {\bf B37} (1988) 3557. 
\bibitem{SC}
G. Semerjian and L.F. Cugliandolo, J. Phys. {\bf A35} (2002) 4837.  
\bibitem{NT}
T. Nagao and T. Tanaka,  J. Phys. {\bf A40} (2007) 4973.
\bibitem{KRKK}
D.-H. Kim, G.J. Rodgers, B. Kahng and D. Kim, Phys. Rev. {\bf E71} 
(2005) 056115.
\bibitem{BM}
G. Biroli and R. Monasson, J. Phys. {\bf A32} (1999) L255. 
\bibitem{BS}
J.N. Bandyopadhyay and S. Jalan, Phys. Rev. {\bf E76} (2007) 026109. 
\bibitem{CLV}
F. Chung, L. Lu and V. Vu, Ann. Comb. {\bf 7} (2003) 21. 

\end{thebibliography}
\end{document}